\documentclass[twocolumn, showpacs,preprintnumbers,amsmath,amssymb]{revtex4}

\usepackage{graphicx}

\begin{document}

\addtolength{\textheight}{1.2cm}
\addtolength{\topmargin}{-0.5cm}

\newcommand{\etal} {{\it et al.}}

\title{Hyperfine-induced decoherence in triangular spin-cluster qubits}

\author{Filippo Troiani$^1$}
\author{Dimitrije Stepanenko$^2$}
\author{Daniel Loss$^2$}

\affiliation{$^1$S3, Istituto Nanoscienze-CNR, via G. Campi 213/A, I-41100 Modena, Italy}
\affiliation{$^2$Department of Physics, University of Basel, Klingbergstrasse 82, CH-4056 Basel, Switzerland}

\date{\today}

\begin{abstract}

We investigate hyperfine-induced decoherence in a triangular spin-cluster for different
qubit encodings. 
Electrically controllable eigenstates of spin chirality ($C_z$) show decoherence times that approach milliseconds, two orders of magnitude longer than those estimated for the eigenstates of the total spin projection ($S_z$) and of the partial spin sum ($S_{12}$). 
The robustness of chirality is due to its decoupling from both the total- and individual-spin components in the cluster. This results in a suppression of the effective interaction between $C_z$ and the nuclear spin bath.

\end{abstract}

\pacs{03.65.Yz,75.50.Xx,03.67.Lx}

\maketitle

{\it Introduction --- }
Molecular nanomagnets represent a varied class of spin clusters, whose physical 
properties can be extensively engineered by chemical synthesis \cite{gatteschi}.
This makes them candidate systems for the implementation of spin-cluster qubits 
\cite{Meier03,Troiani05,Lehmann}.
While most of the attention has been so far focused on the use of the total-spin 
projection ($S_z$) as a computational degree of freedom (DOF), it has been recently 
realized that 
alternative encodings would enable the use of electric - rather than magnetic - 
fields for the qubit manipulation \cite{Trif2008}. 
In particular, transitions between states of opposite spin chirality 
[$C_z = (4/\sqrt{3}) {\bf s}_1 \cdot {\bf s}_2 \times {\bf s}_3 $] can be induced 
in antiferromagnetic triangles with Dzyaloshinskii-Moriya interaction. 
Spin-electric coupling constants compatible with ns gating times $\tau_g$ have 
been predicted by effective models \cite{Trif2008,Trif2010} and microscopic ab 
initio calculations \cite{Fhokrul}.
Further investigation is indeed required in order identify specific molecular 
nanomagnets with large spin-electric coupling, or to enhance such
coupling by introducing suitable chemical substitutions in existing systems \cite{Baadji}. 

In order to assess the suitability of spin chirality for applications in 
quantum-information processing, its $ \tau_g $ has to be contrasted with a 
characteristic decoherence time $ \tau_d $. 
At low temperatures, quantum coherence in molecular nanomagnets is limited by the 
coupling to the nuclear spin environment, with typical values of $ \tau_d $ in the 
microsecond range \cite{Ardavan2007,Bertaina2008,Schlegel2008}.
All the existing literature is however concerned with linear superpositions of 
different $S_z$ eigenstates. 
Here we theoretically investigate the dependence of hyperfine-induced decoherence 
on the qubit encoding within a prototypical spin-cluster qubit, consisting of 
an antiferromagnetic spin triangle. 
In particular, we consider three different DOF, namely $S_z$, $C_z$, 
and the partial spin sum $S_{12}$ (${\bf S}_{12} = {\bf s}_1 + {\bf s}_2 $),
whose value - like that of $C_z$ - can be controlled through the spin-electric 
coupling.
Since the optimal candidate system has not been identified yet, we refer
here to a prototypical molecular spin-cluster qubit, with a typical   
electron-spin Hamiltonian \cite{Choi2006} and bath of nuclear spins 
\cite{Schlegel11}.
While the quantities of interest might to some extent vary from one molecular 
nanomagnet to another, 
the hyperfine-induced decoherence presents striking differences in the three 
considered DOF, that are not expected to depend on the specific features of the 
spin-cluster qubit.

\begin{figure}[ptb]
\begin{center}
\includegraphics[width=8.5cm]{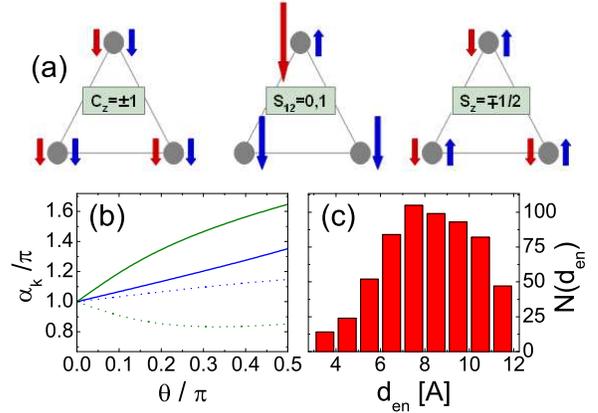}
\end{center}
\caption{(color online) (a) Schematics of the local spin projections $ \langle s_{z,i}
\rangle $ in the spin triangle, corresponding to the $ |0\rangle $ (red) and $ |1\rangle $ (blue) states in the three considered qubit encodings. 
All the states have $S=1/2$; besides, 
$ \langle k | S_z | k \rangle_{C_z} = \langle k | S_z | k \rangle_{S_{12}} = -1/2 $
and
$ \langle k | C_z | k \rangle_{S_z} = 1 $, 
with $ k=0,1$.
(b) Angle $ \alpha_k $ between the vector $ \langle {\bf S} \rangle $ and 
$ \hat{\bf z} $ for the eigenstates $ | 0 \rangle_{C_z}^\theta $ (blue) and 
$ | 1 \rangle_{C_z}^\theta $ (green), for $ \Delta / g\mu_B = 0.5 $ (solid lines) 
and $2.0$ (dotted). 
(c) Statistical distribution of the distances $d_{en}$ between the $N_e=3$ electron 
and the $N_n = 200$ nuclear spins with randomly generated positions.}
\label{fig1}
\end{figure}
{\it Qubit encodings in the spin triangle ---}
We consider a triangle of $s=1/2$ spins, with dominant antiferromagnetic coupling and 
Zeeman interaction: 
\begin{equation}
H_0 = J \sum_{i=1}^3 {\bf s}_i \cdot {\bf s}_{i+1} + g \mu_B {\bf B} \cdot {\bf S} . 
\end{equation}
An additional term $H_1$ determines the expression of the lowest eigenstates 
$ | 0 \rangle $ and $ | 1 \rangle $, belonging to the ground state $S=1/2$ quadruplet. 
As discussed in the following, the robustness of the spin-cluster qubit with respect
to hyperfine-induced decoherence strongly depends on the distinguishability between 
$ | 0 \rangle $ and $ | 1 \rangle $ in terms not only of total spin 
orientation, but also of spin texture.
Hereafter, we thus discuss these features in some detail in two relevant cases:
\begin{eqnarray}
H_1^{C_z}    & = & D \hat{\bf z}\cdot \sum_{i=1}^3 {\bf s}_i \times {\bf s}_{i+1} ,\\
H_1^{S_{12}} & = & (J_{12}-J)\, {\bf s}_1 \cdot {\bf s}_2 .
\label{h1}
\end{eqnarray}
The term $ H_1^{C_z} $ accounts for the Dzyaloshinskii-Moriya interaction in a spin 
triangle with, e.g., $D_{3h}$ symmetry \cite{Choi2006}. 
For $ H_e = H_0 + H_1^{C_z} $, the four lowest eigenstates can be labelled after 
the value of the spin chirality 
$ C_z = (4/\sqrt{3}) {\bf s}_1 \cdot {\bf s}_2 \times {\bf s}_3 $, 
and the Dzyaloshinskii-Moriya term can be rephrased as:
$ H_1^{C_z} = \Delta C_z S_z $, with $ \Delta = D \sqrt{3} $ \cite{Trif2008}.
In particular, if the magnetic field is oriented parallel to the principal axis of the 
molecule ($ {\bf B} \parallel \hat{\bf z} $), the eigenstates $ | C_z, S_z \rangle $ 
read:
$ | \pm 1, +1/2 \rangle = ( |\downarrow\uparrow\uparrow\rangle + e^{\pm i 2\pi /3} |\uparrow\downarrow \uparrow\rangle + e^{\mp i 2\pi /3}  |\uparrow\uparrow\downarrow\rangle ) / \sqrt{3} $ 
and
$ | \pm 1, -1/2 \rangle = \sigma_x^1\sigma_x^2\sigma_x^3  | \pm 1, +1/2 \rangle $,
where $\sigma_x^i$ is the Pauli operator acting on ${\bf s}_i$.
Both $S_z$ and $C_z$ commute with the electron-spin Hamiltonian $ H_e $, 
which makes them suitable as computational DOF. 
In the first case, the logical states are:
\begin{equation*}
| 0 \rangle_{S_z} \!\!\! = | S_z\!\!\! =\!\! -1/2 ; C_z\!\! =\!\! + 1 \rangle ,\
| 1 \rangle_{S_z} \!\!\! = | S_z\!\!\! =\!\! +1/2 ; C_z\!\! =\!\! + 1 \rangle .
\end{equation*}
The expectation values of ${\bf s}_i$ are oriented along the magnetic field,
are identical for the three spins, and change sign with the qubit state 
[Fig. \ref{fig1} (a)]: 
\begin{equation}
\langle 1 | s_{z,i} | 1 \rangle_{S_z} = - \langle 0 | s_{z,i} | 0 \rangle_{S_z} = 1/6 .
\label{evsz}
\end{equation} 
If the computational DOF is identified with spin chirality, the 
logical states are instead: 
\begin{equation*}
|0\rangle_{C_z} \!\!\! = |C_z\!\!\! =\!\! +1; S_z\!\! =\!\! -1/2\rangle, \
|1\rangle_{C_z} \!\!\! = |C_z\!\!\! =\!\! -1; S_z\!\! =\!\! -1/2\rangle ,
\end{equation*}
and the expectation values of the three spins are independent on the 
qubit state [Fig. \ref{fig1} (a)]:
\begin{equation} 
\langle 1 | s_{z,i} | 1 \rangle_{C_z} = \langle 0 | s_{z,i} | 0 \rangle_{C_z} = - 1/6 .
\label{evcz}
\end{equation}
\begin{figure}[ptb]
\begin{center}
\includegraphics[width=8.5cm]{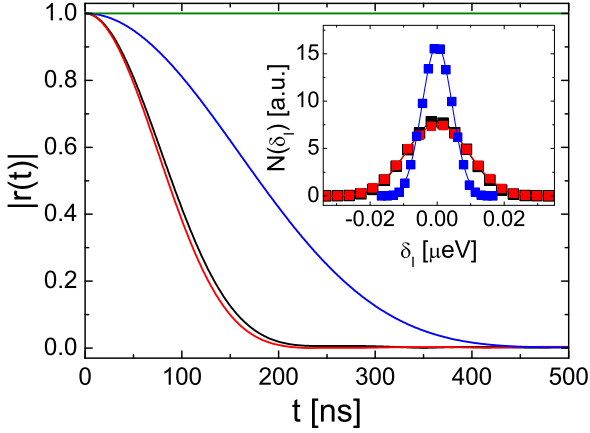}
\end{center}
\caption{(color online) Time dependence of the decoherence factor $r$ for the three
qubit encodings: $S_z$ (black), $S_{12}$ (red), and $C_z$ (green for $\theta 
= 0 $ and blue for $ \theta = \pi / 8$). 
The curves are averaged over $N_\mathcal{I} = 5 \times 10^4$ randomly generated initial 
states $ | \mathcal{I} \rangle $ of the nuclear bath. 
Inset: Statistical distribution (squares) of the parameter $ \delta_\mathcal{I} $, 
and corresponding Gaussian fits (solid lines); same convention as above for the 
colors. }
\label{fig2}
\end{figure}
Such condition is however not general. In fact, if the applied magnetic field is tilted 
with respect to the $z$ axis, 
$ {\bf B} = B ( \sin\theta \hat{\bf x} + \cos\theta \hat{\bf z} ) $, 
$C_z$ is still a good quantum number, but
$ \langle k | {\bf s}_i | k \rangle_{C_z}^\theta $ (with $k=0,1$) 
are always oriented along $ {\bf B}_k' = ( B_x , 0 , B_z \pm \Delta / g\mu_B ) $.
Eigenstates of opposite chirality are thus characterized by different 
orientations of the spin expectation values [see Fig. \ref{fig1}(b)]:
\begin{equation}
\langle k | s_{x,i} | k \rangle_{C_z}^\theta = \sin\alpha_k / 6 , \ 
\langle k | s_{z,i} | k \rangle_{C_z}^\theta = \cos\alpha_k / 6 ,
\label{evtheta}
\end{equation}
where 
$ \alpha_k = \arctan\left[ 
\frac{\chi B\sin\theta}{B\cos\theta + (-1)^k \Delta / g \mu_B}\right] 
+ \pi $,
$ 0 \le \arctan \le \pi $
and
$ \chi = \pm 1 $ 
for $ \Delta \gtrless Bg\mu_B $.

If no Dzyaloshinskii-Moriya interaction is present and one exchange coupling 
differs from the other two, the term $ H_1^{C_z} $ is replaced by $ H_1^{S_{12}} $ 
(Eq. \ref{h1}).
For $ H_e = H_0 + H_1^{S_{12}} $, the four lowest eigenstates can be labelled after 
the partial sum of the first two spins, rather than the spin chirality:
$ | S_{12}, S_z \rangle $, where 
$S_{12}=0,1$.
Their expressions read:
$ | 0, +1/2 \rangle = ( |\uparrow\downarrow\uparrow\rangle - |\downarrow\uparrow\uparrow\rangle ) / \sqrt{2} $,
$ | 1, +1/2 \rangle = ( |\uparrow\downarrow\uparrow\rangle + |\downarrow\uparrow\uparrow\rangle - 2 |\uparrow\uparrow\downarrow\rangle ) / \sqrt{6} $,
while
$ | S_{12}, -1/2 \rangle = \sigma_x^1\sigma_x^2\sigma_x^3 | S_{12}, +1/2 \rangle $.
Choosing $S_{12}$ as the computational DOF, one has: 
\begin{equation*}
|0\rangle_{S_{12}}\!\!\! = | S_{12}=0; S_z\!\!=\!\!-1/2\rangle , \
|1\rangle_{S_{12}}\!\!\! = | S_{12}=1; S_z\!\!=\!\!-1/2\rangle .
\end{equation*}
As far as the spin expectation values are concerned, $S_{12}$ represents an 
intermediate case between $S_z$ and $C_z$.
The qubit states have in fact identical values for the total spin,
$ \langle 0 | {\bf S} | 0 \rangle_{S_{12}} = 
  \langle 1 | {\bf S} | 1 \rangle_{S_{12}} $,
like $C_z$,
but they strongly differ in terms of spin texture, like $S_z$ [Fig. \ref{fig1}(a)]
\begin{subequations}
\begin{eqnarray}
\langle 0 | s_{z,i=1,2} | 0 \rangle_{S_{12}} = 0 , \
\langle 0 | s_{z,3} | 0 \rangle_{S_{12}} = -1/2\ \ \\
\langle 1 | s_{z,i=1,2} | 1 \rangle_{S_{12}} = 
-1/3, \ \langle 1 | s_{z,3} | 1 \rangle_{S_{12}} = 1/6 .
\end{eqnarray}
\end{subequations}

{\it Nuclear spin and hyperfine interactions ---}  
The decoherence of the spin-cluster qubit is
investigated by simulating the coupled dynamics of electron and
nuclear spins, induced by the Hamiltonian $ H = H_e + H_n + H_{en} $.  
The qubit and the nuclear environment are initialized respectively in the linear superposition 
$ | \psi_e (0) \rangle = \frac{1}{\sqrt{2}} ( |0\rangle + |1\rangle ) $
and in the mixed state  
$ \rho_n (0) =
\sum_\mathcal{I} P_\mathcal{I} | \mathcal{I} \rangle\langle
\mathcal{I} | $.  
Here, the expressions of $ |0\rangle $ and $ |1\rangle $ depend on $H_1$
(see above), while 
$ | \mathcal{I} \rangle = |
m^\mathcal{I}_1 , \dots , m^\mathcal{I}_{N_n} \rangle $  and $
m^\mathcal{I}_i $ are the projections along the magnetic field
direction of the $N_n$ nuclear spins.  
In the pure-dephasing regime,
each state 
$ | \Psi_{\mathcal{I}} (0) \rangle = \frac{1}{\sqrt{2}} ( |0\rangle + |1\rangle ) 
\otimes | \mathcal{I} \rangle $, 
evolves into: $ | \Psi_{\mathcal{I}} (t) \rangle =
\frac{1}{\sqrt{2}} ( |0, \mathcal{I}_0 \rangle +  |1,
\mathcal{I}_1\rangle ) $, where $ | \mathcal{I}_0 \rangle $ ($ |
\mathcal{I}_1 \rangle $) can be regarded as  the state of the nuclear
bath conditioned upon the qubit being in the $ |0\rangle $  ($
|1\rangle $) state.  The degree of coherence in the reduced density
matrix of the qubit,  
$ \rho_e = {\rm Tr}_n \{ \sum_{\mathcal{I}} P_\mathcal{I} 
|\Psi_{\mathcal{I}} (t)\rangle\langle \Psi_{\mathcal{I}} (t)| \} $, 
is given by the so-called decoherence factor:
$ r (t) = \sum_\mathcal{I} P_\mathcal{I} r_\mathcal{I} (t) $,
with
$ r_\mathcal{I} (t) = \langle \mathcal{I}_1
(t) | \mathcal{I}_0 (t) \rangle $ 
and 
$ \langle 0 | \rho_e | 1 \rangle =
r_\mathcal{I} / 2 $.  

The nuclear spin bath we consider consists of $ N_n = 200 $ hydrogens
($I = 1/2 $),  whose positions $ {\bf r}_p^n $ are randomly generated
so as to reproduce typical values of the spin density and the
electron-nuclear distances  $ d_{en} = |{\bf r}_i^e - {\bf r}_p^n| $,
where ${\bf r}_i^e$ are the positions of electron spins [Fig. \ref{fig1}
(c)]  \cite{nuclearbath1}.  The nuclear-spin Hamiltonian $H_n$ includes
Zeeman and dipole-dipole terms: $ H_n = \hat{\bf B} \cdot \sum_{p}
\omega_p {\bf I}_p +  D_{nn} \sum_{p<q} [ {\bf I}_p \cdot {\bf I}_q  -
  3 ({\bf I}_p \cdot \hat{\bf r}_{pq}) ({\bf I}_q \cdot \hat{\bf
    r}_{pq}) ] /  r_{pq}^3 $, where $ D_{nn} = ( \mu_0 / 4\pi )
\mu_n^2 \gamma_I^2 $ and  $ {\bf r}_{pq} = {\bf r}_p^n - {\bf r}_q^n
$.  Electron and nuclear spins are coupled by dipole-dipole and
contact interactions: $ H_{en} =   D_{en} \sum_i\sum_p [ {\bf s}_i
  \cdot {\bf I}_p  - 3 ({\bf s}_i \cdot \hat{\bf r}_{ip}) ({\bf I}_p
  \cdot \hat{\bf r}_{ip}) ] /  r_{ip}^3 + \sum_i a_i {\bf s}_i \cdot
{\bf I}_{q(i)} $,  where $ D_{en} = ( \mu_0 / 4\pi ) \mu_n\mu_e
\gamma_I\gamma_e $ and  $ {\bf r}_{ip} = {\bf r}_i^e - {\bf r}_q^n $.
The contact terms $a_i$, whose effect will be considered in the final part
of the paper, couples electron and nuclear spins belonging to the
same magnetic center. 

The dephasing arises from the qubit-state dependent dynamics of the
nuclear bath, generated by an effective Hamiltonian $ \mathcal{H} $.
We derive $ \mathcal{H} $ from the above specified 
$ H = H_e + H_n + H_{en} $
in a two-step procedure \cite{troiani08,Szallas2010}.  
We first project
the single-electron spin operators $ s_{\alpha ,i} $  onto the $S=1/2$
quadruplet:  $ \mathcal{P}_{S=1/2} s_p^\alpha
\mathcal{P}_{S=1/2} =  \sum_{i,j=0}^3 \langle i | s_p^\alpha | j
\rangle   \sigma_{ij} $, where $ \sigma_{ij} = | i \rangle\langle j |
$ and $ | i \rangle $  are the eigenstates of $ H_e $.  We then apply
a Schrieffer-Wolff transformation, that removes from the Hamiltonian
the terms that are off-diagonal in the basis of electron-spin
eigenstates $ | i\rangle $  \cite{wang06,coish}, 
and finally neglect energy non-conserving terms
(secular approximation). The resulting Hamiltonian
reads: $ \mathcal{H} = \sum_{k=0,1} | k \rangle\langle k | \otimes
\mathcal{H}_k $,  where 
\begin{equation}
\mathcal{H}_k = \sum_{p=1}^{N_n} \omega^k_p I_p^{z'} + 
\sum_{p \neq q} ( A^k_{pq} I^{z'}_p I^{z'}_q + B^k_{pq} I^+_p I^-_q )
\end{equation}
and $\hat{\bf z}' \!\!\! \equiv \!\!\! {\bf B} / B $. 
Differences between $ \mathcal{H}_0 $ and $ \mathcal{H}_1 $ result
from the hyperfine  interactions, and are responsible for the qubit
decoherence, being $ r_\mathcal{I} (t) =  \langle \mathcal{I} | \exp\{
i \mathcal{H}_1 t /\hbar \}  \exp\{ -i \mathcal{H}_0 t /\hbar \} |
\mathcal{I} \rangle $.  In particular, the quantities $ ( \omega^0_p -
\omega^1_p ) $ are linear in  $ H_{en} $, and essentially result from
differences in the magnetic field induced by the  nuclear spin $ {\bf
  I}_p $ at the electron-spin positions ${\bf r}_i^e$ (see below).
The terms $ ( A^0_{pq} - A^1_{pq} ) $ and $ ( B^0_{pq} - B^1_{pq}
) $ are instead  quadratic in $ H_{en} $, and result from the
qubit-state dependence of the couplings  between pairs of nuclei,
mediated by virtual transitions of the electron spins.  The time
evolution of the nuclear states $ | \mathcal{I}_k \rangle $ is
computed  within the pair-correlation approximation, where the nuclear
dynamics is traced back  to independent flip-flop transitions between
pairs of nuclear spins \cite{wang06,Yang2008}.

\begin{figure}[ptb]
\begin{center}
\includegraphics[width=8.5cm]{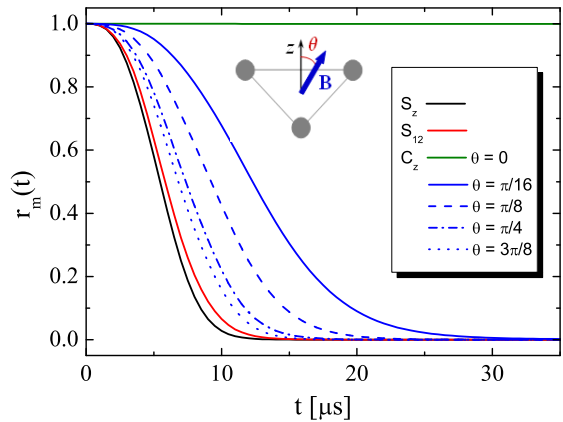}
\end{center}
\caption{(color online) Time evolution of the decoherence factor $ r_m $ in the cases of
the $S_z$ (black) and $S_{12}$ (red) DOF. The case of chirality is displayed for $\theta =0$ (green) and for finite values of the tilting angle (blue). 
All curves are averaged over $10^2$ randomly generated initial conditions 
$ | \mathcal{I} \rangle $; the spin Hamiltonian parameters are: $ \Delta = 1 \,$K, 
$ B = 1\,$T.
}
\label{fig3}
\end{figure}
{\it Hyperfine-induced decoherence --- }  The fastest contribution to
dephasing in the spin-cluster qubit is related to  inhomogeneous
broadening, and typically takes place on time scales that are much
shorter than those characterizing the dynamics of the nuclear bath  ($
\tau_{n} \sim \hbar / | B_{pq}^k |\sim 10^2\,\mu $s).  Such
contribution results from the renormalization of the energy gap
between the states  $ |0\rangle $ and $ | 1 \rangle $ induced by the
hyperfine interactions:  $ \delta_\mathcal{I} = \sum_{k=0,1} (-1)^k
\langle k , \mathcal{I} | \mathcal{H} | k , \mathcal{I} \rangle \simeq
\sum_p (\omega_p^0 - \omega_p^1) m_p^\mathcal{I} $.  Being the nuclear
spin bath initially in a mixture of states $ | \mathcal{I} \rangle $,
the decoherence factor evolves as: $ r(t \ll \tau_{n}) \simeq
e^{-i(E_0-E_1)t} \sum_\mathcal{I} P_\mathcal{I}
e^{-i\delta_\mathcal{I}t} $,  while  $ | \mathcal{I}_k (t \ll
\tau_{n}) \rangle \simeq | \mathcal{I} \rangle $.  In first order in
$H_{en}$, $ \delta_\mathcal{I} $ can be regarded as a function
of the Overhauser field at the electron-spin sites:
\begin{equation}\label{deltaI}
\delta_\mathcal{I} \simeq \mu_B g \sum_i {\bf B}_{hf}^\mathcal{I} ({\bf r}^e_i) 
\cdot [ \langle 0 | {\bf s}_i | 0 \rangle - \langle 1 | {\bf s}_i | 1 \rangle ] ,
\end{equation}
where $ {\bf B}_{hf}^\mathcal{I} ({\bf r}^e_i) =  D_{en} \sum_p
m^\mathcal{I}_p [ \hat{\bf z}' -3 (\hat{\bf z}'\cdot\hat{\bf
    r}_{ip})\hat{\bf r}_{ip} ] / r_{ip}^3 $.  In the case of the $S_z$
qubit (see Eq. \ref{evsz}), $ \delta^{S_z}_\mathcal{I} \simeq -(\mu_B
g/3) \sum_i B_{hf,z'}^\mathcal{I} ({\bf r}^e_i) $.  The statistical
distribution $ N(\delta_\mathcal{I}^{S_z}) $ is reported in the inset
of Fig. \ref{fig2} (black squares) for $5 \times 10^4 $ initial
nuclear states $| \mathcal{I}\rangle $, randomly generated from a flat
probability distribution $ P_\mathcal{I} = 1/2^{N_n} $.  $
N(\delta_\mathcal{I}^{S_z}) $ is well fitted by a Gaussian function
(solid line), with $ \sigma_{S_z} = 9.0\, $neV.  Correspondingly, the
decay of $ | r(t) | $ (black line in Fig. \ref{fig2}) is approximately
Gaussian, and its characteristic time scale is $10^2\,$ns.  In the
case of the $S_{12}$ qubit, the three electron spins are no longer
equivalent:  $ \delta^{S_{12}}_\mathcal{I} \simeq -(\mu_B g/3)
[2B_{hf,z'}^\mathcal{I} ({\bf r}^e_3) - B_{hf,z'}^\mathcal{I} ({\bf
    r}^e_1) - B_{hf,z'}^\mathcal{I} ({\bf r}^e_2)]$.  However,  the
statistical distribution of $\delta_\mathcal{I}^{S_{12}}$  strongly
resembles that of $S_z$ (see the red squares in the figure inset, and
the  Gaussian fit with $\sigma_{S_{12}}=9.4\,$neV),  and so does the
time evolution of the decoherence factor (red curve in the main
panel).  In fact, since the distances $ d_{ee} $ between electron
spins are larger than the smallest $ d_{en} $ [see Fig. \ref{fig1}(c)]
\cite{nuclearbath1}, the spatial fluctuations of the Overhauser field
within the spin triangle are comparable to its average value.  In spin
clusters with larger $ d_{en} / d_{ee} $ ratios (not shown here),
spatial  fluctuations of $ {\bf B}_{hf} ({\bf r}) $ are relatively
small. As a result, $ \delta^{S_{12}}_\mathcal{I} \ll
\delta^{S_{z}}_\mathcal{I}$, and the $S_{12}$ qubit  is less affected
by inhomogeneous broadening than $S_{z}$.

In the case of the $C_z$ qubit and for ${\bf B} \parallel \hat{\bf z}
$,  the Overhauser field does not renormalize the energy difference
between the states $ |0\rangle $ and $ |1\rangle $,  that have identical
expectation values for all single-spin projections (Eqs. \ref{evcz},
\ref{deltaI}).
The leading contribution to $\delta_\mathcal{I}^{C_z}$ is given by
terms that are  second order in the hyperfine Hamiltonian, $
\delta_\mathcal{I}^{C_z} = \sum_{p \neq q} (A^0_{pq}-A^1_{pq})
m_p^\mathcal{I} m_q^\mathcal{I} $, and its modulus is here 5 orders of
magnitude smaller than that of  $ \delta_\mathcal{I}^{S_z} $ and $
\delta_\mathcal{I}^{S_{12}} $.  Correspondingly, no inhomogeneous
broadening occurs in the considered time scale (green curve).  For a
tilted magnetic field ($ \theta \neq 0$), states of opposite chirality
have different expectation values $ \langle {\bf s}_i \rangle $ (see
Eq. \ref{evtheta}), and thus couple differently to the Overhauser
field.  The leading contribution to the renormalization of the energy
difference reads: $ \delta_\mathcal{I}^{C_z} ( \theta ) \simeq  (\mu_B
g /6) 
\sum_{k=0}^1 (-1)^k (\sin\alpha_k B^\mathcal{I}_{hf,x'} +
\cos\alpha_k B^\mathcal{I}_{hf,z'} )  $, where ${\bf x}' \perp {\bf
  z}'$ and lies in the $xz$ plane.  The statistical distribution of $
\delta_\mathcal{I}^{C_z} (\theta = \pi /8) $ and the  resulting qubit
dephasing are reported in Fig. \ref{fig2} ($\sigma_{C_z}=4.5\,$neV,
blue curve).

\begin{figure}[ptb]
\begin{center}
\includegraphics[width=8.5cm]{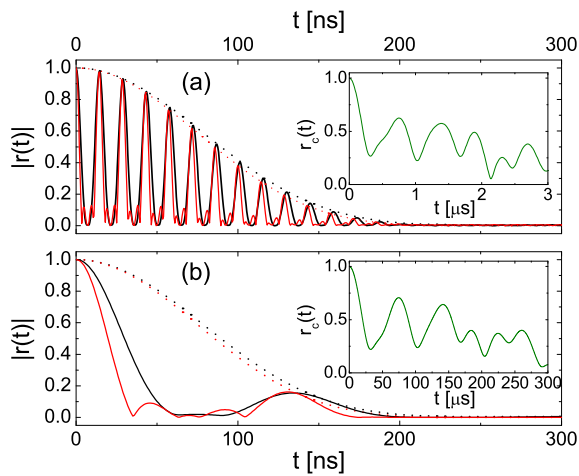}
\end{center}
\caption{(color online)
Time dependence of the decoherence factor in the presence of three additional nuclear
spins ($N_n = 203$) localized at the electron-spin positions $ {\bf r}^e_i $ and coupled to the respective electron spins with a contact coupling $a_{p}=10\,$mK (panel a), or 
$a_{p}=1\,$mK (panel b). 
The solid curves correspond to the cases $S_z$ (black), $S_{12}$ (red) and $C_z$ 
(figure insets). The dotted lines represent the time dependence of $r_m$ in the absence of 
the three nuclei with contact couplings and correspond to those displayed in Fig. 
\ref{fig2}.}
\label{fig4}
\end{figure}
Nuclear spin evolution in the field of electron spins creates
entanglement in the initially separable state $|\Psi_\mathcal{I} (0)\rangle$, 
and causes additional decoherence on the time scale of nuclear dynamics,
$\tau_n$.  In order to single out this contribution, we compute the
function: $ r_{m} (t) = \sum_\mathcal{I} P_\mathcal{I} | r_\mathcal{I}
(t) | $.  In the case of the $S_z$ qubit, electron-nuclear
correlations result in a decay of  $r_m$ in the $\mu$s time scale
(Fig. \ref{fig4}, black curve).  The decay is induced by the interplay
of the dipolar interactions between the nuclei and of the term $
\sum_p (\omega_p^0 - \omega_p^1) I_p^z $, whose expectation value
gives $ \delta_\mathcal{I}^{S_z} $ \cite{nota}.  
A similar time dependence for
$r_m$ is obtained in the case of the $S_{12}$ qubit (red curve). Here,
the same terms in the effective Hamiltonian $ \mathcal{H} $  dominate,
and have similar expectation values: $ \delta_\mathcal{I}^{S_z} \simeq
\delta_\mathcal{I}^{S_{12}} $ (see the inset of Fig. \ref{fig2}).
This quantity ($\delta_\mathcal{I}^{C_z}$) is about 5 orders of
magnitude smaller for the $C_z$ qubit, if $ {\bf B} \parallel \hat{\bf
  z} $.  As a result, the dynamics of the nuclear bath is largely
independent on the qubit  state in the considered time range, and no
appreciable decoherence takes place  (green curve).  This is no longer
the case if the magnetic field is not aligned with the principal  axis
of the spin triangle: in fact, the decoherence time rapidly decreases
as $ \delta_\mathcal{I}^{C_z} ( \theta ) $ increases with the tilting
angle $\theta$ (blue curves). 

We finally investigate the possible contribution to decoherence of the 
contact terms. Such contribution is qualitatively different from that of 
the dipolar interactions, for it results from the relatively strong 
coupling with the electron spins of few ($ N_n^c \sim N_e \ll N_n $) 
nuclei.  
Here, the $ N_n^c = N_e = 3 $ additional nuclear spins are localized at 
the electron spin sites $ {\bf r}_{i}^e $, and are assumed for simplicity
identical to the remaining $ N_n^d = N_n - N_n^c = 200$ nuclei. 
The inequivalence between the $ N_n^c $ and $ N_n^d $ nuclear spins, 
resulting from strong coupling of the former ones with the electron spins, 
warrants the factorization of the decoherence factor:  
$ r (t) = r_c(t) \, r_d(t) $ \cite{inequivalent}.  
The time evolution of $r_c(t)$ is reported in Fig. \ref{fig4}, for  $
a_p = 1 \,$mK and $a_p=10 \,$mK [panels (a) and (b), respectively].
In the case of $S_z$ (black curve) and $S_{12}$ (red), $r_c$ is
responsible for the fast oscillations, while the decay is due to 
$r_d$ (dotted lines). Oscillations of the decoherence factor caused by 
the contact interactions are also present in the case of $C_z$ qubit 
(figure insets). These take place on a time scale which is much longer 
than that of $S_z$ and $S_{12}$, but much shorter than the one that 
characterizes the contribution to the decoherence of $C_z$ due to dipolar 
coupling. The chirality qubit also presents a different dependence on the 
contact coupling constant $a_i$ with respect to $S_z$ and $S_{12}$.
A comparison between the two panels shows in fact that the characteristic 
time scale of the oscillations in $r_c$ is $ \tau_d^c \sim \hbar / a_p$ 
for $S_z$ and $S_{12}$, and $ \tau_d^c \sim \hbar \ \delta_{ij} / a_p^2 $ 
for $ C_z $,  where  $ \delta_{ij} \sim \min \{ \Delta , g\mu_BB \} $ is 
the smallest difference  between eigenvalues of $H_e$. 
Like in the case of dipolar interactions discussed above, the leading 
contributions of contact interaction to $ \mathcal{H}$ are thus quadratic 
in the hyperfine Hamiltonian for $C_z$, and linear for the other two DOF.

In conclusion, we have shown that the nuclear-induced decoherence in a 
prototypical spin triangle strongly depends on the qubit 
encoding. In particular, the decoherence time of the chirality qubit approaches 
the ms range, i.e. several orders of magnitude larger than the gating times
predicted for the manipulation by means of electric fields.  
Such robustness results from the decoupling of $C_z$ from both the total
spin orientation and the spin texture, preserved by the alignment of
the magnetic fields in the direction normal to the spin triangle.
The eigenstates of $S_{12}$ are characterized instead by different 
spin textures, and thus couple differently to the nuclear spin bath. 
This results in decoherence times comparable to those of the total spin 
projection $S_z$.  
Larger decoherence times for $S_{12}$ could be obtained in spin clusters 
where the distance between electron spins is much smaller than that between
electron and nuclear spins. 
Finally, the presence of a strong contact interaction between nuclear and 
electron spins introduces an oscillationg behavior in the decoherence factor 
of all the DOF, with characteristic time scales that are $10^2$ times larger 
for spin chirality than for $S_z$ and $S_{12}$.

We acknowledge financial support by PRIN of the Italian MIUR, by the Swiss NF,
and by FP7-ICT project "ELFOS".


\begin{thebibliography}{21}
\expandafter\ifx\csname natexlab\endcsname\relax\def\natexlab#1{#1}\fi
\expandafter\ifx\csname bibnamefont\endcsname\relax
  \def\bibnamefont#1{#1}\fi
\expandafter\ifx\csname bibfnamefont\endcsname\relax
  \def\bibfnamefont#1{#1}\fi
\expandafter\ifx\csname citenamefont\endcsname\relax
  \def\citenamefont#1{#1}\fi
\expandafter\ifx\csname url\endcsname\relax
  \def\url#1{\texttt{#1}}\fi
\expandafter\ifx\csname urlprefix\endcsname\relax\def\urlprefix{URL }\fi
\providecommand{\bibinfo}[2]{#2}
\providecommand{\eprint}[2][]{\url{#2}}

\bibitem[{\citenamefont{Gatteschi et~al.}(2007)\citenamefont{Gatteschi,
  Sessoli, and Villain}}]{gatteschi}
\bibinfo{author}{\bibfnamefont{D.}~\bibnamefont{Gatteschi}},
  \bibinfo{author}{\bibfnamefont{R.}~\bibnamefont{Sessoli}}, \bibnamefont{and}
  \bibinfo{author}{\bibfnamefont{J.}~\bibnamefont{Villain}},
  \emph{\bibinfo{title}{Molecular nanomagnets}} (\bibinfo{publisher}{Oxford
  University Press}, \bibinfo{year}{2007}).

\bibitem[{\citenamefont{Meier et~al.}(2003)\citenamefont{Meier, Levy, and
  Loss}}]{Meier03}
\bibinfo{author}{\bibfnamefont{F.}~\bibnamefont{Meier}},
  \bibinfo{author}{\bibfnamefont{J.}~\bibnamefont{Levy}}, \bibnamefont{and}
  \bibinfo{author}{\bibfnamefont{D.}~\bibnamefont{Loss}},
  \bibinfo{journal}{Phys. Rev. Lett.} \textbf{\bibinfo{volume}{90}},
  \bibinfo{pages}{047901} (\bibinfo{year}{2003}).

\bibitem[{\citenamefont{Troiani et~al.}(2005)\citenamefont{Troiani, Ghirri,
  Affronte, Carretta, Santini, Amoretti, Piligkos, Timco, and
  Winpenny}}]{Troiani05}
\bibinfo{author}{\bibfnamefont{F.}~\bibnamefont{Troiani}},
  \bibinfo{author}{\bibfnamefont{A.}~\bibnamefont{Ghirri}},
  \bibinfo{author}{\bibfnamefont{M.}~\bibnamefont{Affronte}},
  \bibinfo{author}{\bibfnamefont{S.}~\bibnamefont{Carretta}},
  \bibinfo{author}{\bibfnamefont{P.}~\bibnamefont{Santini}},
  \bibinfo{author}{\bibfnamefont{G.}~\bibnamefont{Amoretti}},
  \bibinfo{author}{\bibfnamefont{S.}~\bibnamefont{Piligkos}},
  \bibinfo{author}{\bibfnamefont{G.}~\bibnamefont{Timco}}, \bibnamefont{and}
  \bibinfo{author}{\bibfnamefont{R.~E.~P.}~\bibnamefont{Winpenny}},
  \bibinfo{journal}{Phys. Rev. Lett.} \textbf{\bibinfo{volume}{94}},
  \bibinfo{pages}{207208} (\bibinfo{year}{2005}).

\bibitem[{\citenamefont{Lehmann et~al.}(2007)\citenamefont{Lehmann,
  Gaita-Ari\~{n}o, Coronado, and Loss}}]{Lehmann}
\bibinfo{author}{\bibfnamefont{J.}~\bibnamefont{Lehmann}},
  \bibinfo{author}{\bibfnamefont{A.}~\bibnamefont{Gaita-Ari\~{n}o}},
  \bibinfo{author}{\bibfnamefont{E.}~\bibnamefont{Coronado}}, \bibnamefont{and}
  \bibinfo{author}{\bibfnamefont{D.}~\bibnamefont{Loss}},
  \bibinfo{journal}{Nature Nanotech.} \textbf{\bibinfo{volume}{2}},
  \bibinfo{pages}{312} (\bibinfo{year}{2007}).

\bibitem[{\citenamefont{Trif et~al.}(2008)\citenamefont{Trif, Troiani,
  Stepanenko, and Loss}}]{Trif2008}
\bibinfo{author}{\bibfnamefont{M.}~\bibnamefont{Trif}},
  \bibinfo{author}{\bibfnamefont{F.}~\bibnamefont{Troiani}},
  \bibinfo{author}{\bibfnamefont{D.}~\bibnamefont{Stepanenko}},
  \bibnamefont{and} \bibinfo{author}{\bibfnamefont{D.}~\bibnamefont{Loss}},
  \bibinfo{journal}{Phys. Rev. Lett.} \textbf{\bibinfo{volume}{101}},
  \bibinfo{pages}{217201} (\bibinfo{year}{2008}).

\bibitem[{\citenamefont{Trif et~al.}(2010)\citenamefont{Trif, Troiani,
  Stepanenko, and Loss}}]{Trif2010}
\bibinfo{author}{\bibfnamefont{M.}~\bibnamefont{Trif}},
  \bibinfo{author}{\bibfnamefont{F.}~\bibnamefont{Troiani}},
  \bibinfo{author}{\bibfnamefont{D.}~\bibnamefont{Stepanenko}},
  \bibnamefont{and} \bibinfo{author}{\bibfnamefont{D.}~\bibnamefont{Loss}},
  \bibinfo{journal}{Phys. Rev. B} \textbf{\bibinfo{volume}{82}},
  \bibinfo{pages}{045429} (\bibinfo{year}{2010}).

\bibitem[{\citenamefont{Islam et~al.}(2010)\citenamefont{Islam, Nossa, Canali,
  and Pederson}}]{Fhokrul}
\bibinfo{author}{\bibfnamefont{M.~F.} \bibnamefont{Islam}},
  \bibinfo{author}{\bibfnamefont{J.~F.} \bibnamefont{Nossa}},
  \bibinfo{author}{\bibfnamefont{C.~M.} \bibnamefont{Canali}},
  \bibnamefont{and} \bibinfo{author}{\bibfnamefont{M.}~\bibnamefont{Pederson}},
  \bibinfo{journal}{Phys. Rev. B} \textbf{\bibinfo{volume}{82}},
  \bibinfo{pages}{155446} (\bibinfo{year}{2010}).

\bibitem[{\citenamefont{Baadji et~al.}(2009)\citenamefont{Baadji, Piacenza,
  Tugsuz, Della~Sala, Maruccio, and Sanvito}}]{Baadji}
\bibinfo{author}{\bibfnamefont{N.}~\bibnamefont{Baadji}},
  \bibinfo{author}{\bibfnamefont{M.}~\bibnamefont{Piacenza}},
  \bibinfo{author}{\bibfnamefont{T.}~\bibnamefont{Tugsuz}},
  \bibinfo{author}{\bibfnamefont{F.}~\bibnamefont{Della~Sala}},
  \bibinfo{author}{\bibfnamefont{G.}~\bibnamefont{Maruccio}}, \bibnamefont{and}
  \bibinfo{author}{\bibfnamefont{S.}~\bibnamefont{Sanvito}},
  \bibinfo{journal}{Nature Mater.} \textbf{\bibinfo{volume}{8}},
  \bibinfo{pages}{813} (\bibinfo{year}{2009}).

\bibitem[{\citenamefont{Ardavan et~al.}(2007)\citenamefont{Ardavan, Rival,
  Morton, Blundell, Tyryshkin, Timco, and Winpenny}}]{Ardavan2007}
\bibinfo{author}{\bibfnamefont{A.}~\bibnamefont{Ardavan}},
  \bibinfo{author}{\bibfnamefont{O.}~\bibnamefont{Rival}},
  \bibinfo{author}{\bibfnamefont{J.}~\bibnamefont{Morton}},
  \bibinfo{author}{\bibfnamefont{S.}~\bibnamefont{Blundell}},
  \bibinfo{author}{\bibfnamefont{A.}~\bibnamefont{Tyryshkin}},
  \bibinfo{author}{\bibfnamefont{G.}~\bibnamefont{Timco}}, \bibnamefont{and}
  \bibinfo{author}{\bibfnamefont{R.~E.~P.}~\bibnamefont{Winpenny}},
  \bibinfo{journal}{Phys. Rev. Lett.} \textbf{\bibinfo{volume}{98}},
  \bibinfo{pages}{057201} (\bibinfo{year}{2007}).

\bibitem[{\citenamefont{Bertaina et~al.}(2008)\citenamefont{Bertaina,
  Gambarelli, Mitra, Tsukerblat, M\"{u}ller, and Barbara}}]{Bertaina2008}
\bibinfo{author}{\bibfnamefont{S.}~\bibnamefont{Bertaina}},
  \bibinfo{author}{\bibfnamefont{S.}~\bibnamefont{Gambarelli}},
  \bibinfo{author}{\bibfnamefont{T.}~\bibnamefont{Mitra}},
  \bibinfo{author}{\bibfnamefont{B.}~\bibnamefont{Tsukerblat}},
  \bibinfo{author}{\bibfnamefont{A.}~\bibnamefont{M\"{u}ller}},
  \bibnamefont{and} \bibinfo{author}{\bibfnamefont{B.}~\bibnamefont{Barbara}},
  \bibinfo{journal}{Nature} \textbf{\bibinfo{volume}{453}},
  \bibinfo{pages}{203} (\bibinfo{year}{2008}).

\bibitem[{\citenamefont{Schlegel et~al.}(2008)\citenamefont{Schlegel, van
  Slageren, Manoli, Brechin, and Dressel}}]{Schlegel2008}
\bibinfo{author}{\bibfnamefont{C.}~\bibnamefont{Schlegel}},
  \bibinfo{author}{\bibfnamefont{J.}~\bibnamefont{van Slageren}},
  \bibinfo{author}{\bibfnamefont{M.}~\bibnamefont{Manoli}},
  \bibinfo{author}{\bibfnamefont{E.~K.} \bibnamefont{Brechin}},
  \bibnamefont{and} \bibinfo{author}{\bibfnamefont{M.}~\bibnamefont{Dressel}},
  \bibinfo{journal}{Phys. Rev. Lett.} \textbf{\bibinfo{volume}{101}},
  \bibinfo{pages}{147203} (\bibinfo{year}{2008}).

\bibitem[{\citenamefont{Choi et~al.}(2006)\citenamefont{Choi, Matsuda, Nojiri,
  Kortz, Hussain, Stowe, Ramsey, and Dalal}}]{Choi2006}
\bibinfo{author}{\bibfnamefont{K.-Y.} \bibnamefont{Choi}},
  \bibinfo{author}{\bibfnamefont{Y.}~\bibnamefont{Matsuda}},
  \bibinfo{author}{\bibfnamefont{H.}~\bibnamefont{Nojiri}},
  \bibinfo{author}{\bibfnamefont{U.}~\bibnamefont{Kortz}},
  \bibinfo{author}{\bibfnamefont{F.}~\bibnamefont{Hussain}},
  \bibinfo{author}{\bibfnamefont{A.}~\bibnamefont{Stowe}},
  \bibinfo{author}{\bibfnamefont{C.}~\bibnamefont{Ramsey}}, \bibnamefont{and}
  \bibinfo{author}{\bibfnamefont{N.~S.}~\bibnamefont{Dalal}},
  \bibinfo{journal}{Phys. Rev. Lett.} \textbf{\bibinfo{volume}{96}},
  \bibinfo{pages}{107202} (\bibinfo{year}{2006}).

\bibitem[{\citenamefont{Schlegel et~al.}(2011)\citenamefont{Schlegel, van
  Slageren, Timco, Winpenny, and Dressel}}]{Schlegel11}
\bibinfo{author}{\bibfnamefont{C.}~\bibnamefont{Schlegel}},
  \bibinfo{author}{\bibfnamefont{J.}~\bibnamefont{van Slageren}},
  \bibinfo{author}{\bibfnamefont{G.}~\bibnamefont{Timco}},
  \bibinfo{author}{\bibfnamefont{R.~E.~P.} \bibnamefont{Winpenny}},
  \bibnamefont{and} \bibinfo{author}{\bibfnamefont{M.}~\bibnamefont{Dressel}},
  \bibinfo{journal}{Phys. Rev. B} \textbf{\bibinfo{volume}{83}},
  \bibinfo{pages}{134407} (\bibinfo{year}{2011}).

\bibitem[{nuc()}]{nuclearbath1}
\bibinfo{note}{The distances between the elecron spins are $ | {\bf r}_i^e -
  {\bf r}_j^e | = 5\, $\AA. The nuclei are distributed in a sphere of radius
  $R=10\,$\AA centred in the origin, with the following constraints: $ | {\bf
  r}_i^n - {\bf r}_j^n | \ge 1.5\, $\AA\ and $ | {\bf r}_i^n - {\bf r}_j^e |
  \ge 3\, $\AA.}

\bibitem[{\citenamefont{Szallas and Troiani}(2010)}]{Szallas2010}
\bibinfo{author}{\bibfnamefont{A.}~\bibnamefont{Szallas}} \bibnamefont{and}
  \bibinfo{author}{\bibfnamefont{F.}~\bibnamefont{Troiani}},
  \bibinfo{journal}{Phys. Rev. B} \textbf{\bibinfo{volume}{82}},
  \bibinfo{pages}{224409} (\bibinfo{year}{2010}).

\bibitem[{\citenamefont{Troiani et~al.}(2008)\citenamefont{Troiani, Bellini,
  and Affronte}}]{troiani08}
\bibinfo{author}{\bibfnamefont{F.}~\bibnamefont{Troiani}},
  \bibinfo{author}{\bibfnamefont{V.}~\bibnamefont{Bellini}}, \bibnamefont{and}
  \bibinfo{author}{\bibfnamefont{M.}~\bibnamefont{Affronte}},
  \bibinfo{journal}{Phys. Rev. B} \textbf{\bibinfo{volume}{77}},
  \bibinfo{pages}{054428} (\bibinfo{year}{2008}).

\bibitem[{\citenamefont{Yao et~al.}(2006)\citenamefont{Yao, Liu, and
  Sham}}]{wang06}
\bibinfo{author}{\bibfnamefont{W.}~\bibnamefont{Yao}},
  \bibinfo{author}{\bibfnamefont{R.-B.} \bibnamefont{Liu}}, \bibnamefont{and}
  \bibinfo{author}{\bibfnamefont{L.~J.} \bibnamefont{Sham}},
  \bibinfo{journal}{Phys. Rev. B} \textbf{\bibinfo{volume}{74}},
  \bibinfo{pages}{195301} (\bibinfo{year}{2006}).

\bibitem[{\citenamefont{Coish et~al.}(2008)\citenamefont{Coish, Fischer, and
  Loss}}]{coish}
\bibinfo{author}{\bibfnamefont{W.~A.} \bibnamefont{Coish}},
  \bibinfo{author}{\bibfnamefont{J.}~\bibnamefont{Fischer}}, \bibnamefont{and}
  \bibinfo{author}{\bibfnamefont{D.}~\bibnamefont{Loss}},
  \bibinfo{journal}{Phys. Rev. B} \textbf{\bibinfo{volume}{77}},
  \bibinfo{pages}{125329} (\bibinfo{year}{2008}).

\bibitem[{\citenamefont{Yang and Liu}(2008)}]{Yang2008}
\bibinfo{author}{\bibfnamefont{W.}~\bibnamefont{Yang}} \bibnamefont{and}
  \bibinfo{author}{\bibfnamefont{R.-B.} \bibnamefont{Liu}},
  \bibinfo{journal}{Phys. Rev. B} \textbf{\bibinfo{volume}{78}},
  \bibinfo{pages}{085315} (\bibinfo{year}{2008}).

\bibitem[{not()}]{nota}
\bibinfo{note}{The time evolution of $r_m$ obtained for $S_z$ with the full
  Hamiltonian $ \mathcal{H} $ is in fact indistinguishable from that obtained
  by keeping in $ \mathcal{H} $ only the terms that are linear in the hyperfine
  couplings.}

\bibitem[{ine()}]{inequivalent}
\bibinfo{note}{The hyperfine field induced by the contact terms is in fact of
  the order of $ 1 \div 10 \,$T. This makes the occurrence of a flip-flop
  transition between an ${\bf I}_p$ with an ${\bf I}_q$ without contact
  interaction highly unlikely, being: $ | \omega_p - \omega_q | \sim a_p \gg
  B_{pq}^k \sim D_{nn} / r_{pq}^3 $.}

\end{thebibliography}
\end{document}